\documentclass[%
 reprint,
superscriptaddress,
 amsmath,amssymb,
 aps,
]{revtex4-2}

\usepackage{graphicx}
\usepackage{dcolumn}
\usepackage{bm}
\usepackage{bbm}
\usepackage{subcaption}
\usepackage{mwe}
\usepackage{verbatim}

\begin{document}

\preprint{APS/123-QED}

\title{Measurement disturbance tradeoffs in three-qubit unsupervised quantum classification}

%


\author{Hector Spencer-Wood}
 \email{h.spencer-wood.1@research.gla.ac.uk}
\affiliation{School of Physics and Astronomy, University of Glasgow, Glasgow G12 8QQ, Scotland}%

\author{John Jeffers}%
\affiliation{Department of Physics, University of Strathclyde, John Anderson Building, 107 Rottenrow, Glasgow, G4 0NG, Scotland}

\author{Sarah Croke}
 \affiliation{School of Physics and Astronomy, University of Glasgow, Glasgow G12 8QQ, Scotland}

\date{\today}
\begin{abstract}
We consider measurement disturbance tradeoffs in quantum machine learning protocols which seek to learn about quantum data. We study the simplest example of a binary classification task, in the unsupervised regime. Specifically, we investigate how a classification of two qubits, that can each be in one of two unknown states, affects our ability to perform a subsequent classification on three qubits when a third is added. Surprisingly, we find a range of strategies in which a non-trivial first classification does not affect the success rate of the second classification. There is, however, a non-trivial measurement disturbance tradeoff between the success rate of the first and second classifications, and we fully characterise this tradeoff analytically.
\end{abstract}

\maketitle


\section{Introduction}

The proliferation of huge datasets in modern science, technology, and society in general has spurred rapid developments in machine learning; a powerful set of techniques which seek to automate the drawing of inferences from data. A recent theoretical development has been to apply ideas from machine learning to the processing of quantum data \cite{dunjko2018Review}, both in a supervised setting \cite{ProgrammableQSD,guta2010quantum,sasaki2001quantum} and an unsupervised setting \cite{UnsupervisedClassification,HayashiQuantChange,sentis2016quantum}, for example, at the output of a quantum communication, sensing, or processing device. In the longer run, as quantum technologies develop further, such techniques may be expected to find use in e.g. characterising quantum channels and devices, including monitoring for malfunctions \cite{HayashiQuantChange,sentis2016quantum}. Indeed, in \cite{HayashiQuantChange,sentis2016quantum}, the problem of determining a quantum change point is addressed. Here, the change point could be the result of some unknown error in a quantum device outputting quantum data. We note that another prominent line of research in quantum machine learning is that of using quantum processing techniques to aid and speed up machine learning when applied to classical data \cite{dunjko2018Review}. We, however, only consider the task of learning about quantum data, which requires rather different techniques.

Quantum data is fundamentally different to classical data, and learning strategies are therefore subject to different, peculiarly quantum limitations, which are not yet well explored. As an example, quantum data famously cannot be cloned \cite{woottersCloning,dieksCloning}, in stark contrast to the classical case. In addition, it is not possible to extract information about a quantum system without causing disturbance \cite{Uncertainty}. Measurement strategies must therefore be carefully chosen and generically (but not always) the globally optimal strategy for any learning task involves waiting until all data has been received and then performing a joint measurement over all systems \cite{UnsupervisedClassification,inductive, GlobalVsLocal1, GlobalVsLocal2,GlobalVsLocal3,GlobalVsLocal4,GlobalVsLocal5,GlobalVsLocal6}.

Such considerations thus pose a problem unique to the quantum case: can we learn about a subset of data without compromising performance on the dataset as a whole? We might expect a measurement-disturbance type tradeoff between performance on the subset and performance on the whole dataset. In this paper we take the first steps towards understanding this tradeoff, studying the simplest case of unsupervised binary classification of qubit states, with three samples. A binary classification task is one in which the aim is to assign each sample provided to one of two possible classes, as accurately as possible. Unsupervised means that there is no labelled training data provided, and the user or algorithm must do as well as possible by comparing the data samples to each other. We give analytically the precise tradeoff between learning about the first two samples provided and learning about all three samples. This case is simple enough to allow analytic results, while rich enough to demonstrate the tradeoff. Surprisingly, for a range of strategies on the first two qubits, it is possible to avoid any reduction in performance on all three.

Our work is related to the problem of sequential observers extracting information about a system \cite{Sequential1,Sequential2,Sequential3,Sequential4}, however, so far, the literature has mostly considered the case in which sequential observers have access to the \emph{same} system. Here, in the learning scenario, we are interested in how measurements on some part of a system (the first two subsystems in the example considered here) affect measurement on the whole. In addition, prior work has considered the supervised learning case, in which a labelled training set is provided and used to induce a function to label test instances. Here it is known that in the limit of many test instances, global measurements over training and test data are not required for optimal performance, and the training data may be measured in advance without access to the test data \cite{inductive}. The unsupervised case is more complicated, as the algorithm seeks to both learn from and classify each instance provided.

In the remainder of this paper we will introduce the unsupervised binary classification problem, illustrate the measurement disturbance effect in the learning scenario, and quantify the tradeoff between learning about two samples and learning about all three.

\section{Background theory}

We begin by noting the background theory and notation that will be used throughout this paper. First, as was alluded to earlier, we will be classifying qubits. A qubit $|\varphi\rangle$ is an element of the two dimensional complex Hilbert space $\mathbb{C}^2$ such that $\langle \varphi| \varphi\rangle = 1$. It will be notationally convenient to consider our qubits as spin-$\frac12$ particles. With this in mind, we can define the computational basis states of our qubits:
\begin{align}
    \begin{aligned}
        |0\rangle &:= \!\left| s=\frac12, m_s=\frac12 \!\right\rangle\!,\\
        |1\rangle &:= \!\left| s=\frac12, m_s=-\frac12 \!\right\rangle\!,
    \end{aligned}
\end{align}
where $s,m_s$ denote the total spin and $z$-component of total spin of the system respectively. Now, the classification problem we will be considering is: given a number of qubits that can be in one of two unknown states $|\varphi_0\rangle,|\varphi_1\rangle$, how well can we assign a label, $|\varphi_0\rangle$ or $|\varphi_1\rangle$, to each of them? Being qubits, $|\varphi_k\rangle$ can be visualised as points on the Bloch sphere \cite{barnett2009quantum}. We can therefore explicitly write them as
\begin{equation}
    |\varphi_k\rangle = \cos \frac{\theta_k}{2}|0\rangle + e^{i\phi_k} \sin \frac{\theta_k}{2}|1\rangle,
\end{equation}
where $k\in\{0,1\},$ and $\theta_k \in [0,\pi],~\phi_k \in [0,2\pi)$ are the polar coordinates of a point on the Bloch sphere. 

We will later see that a quantum classification can be formulated as a quantum measurement. A quantum measurement is mathematically equivalent to a positive operator-valued measure (POVM) \cite{barnett2009quantum}. We therefore define a measurement as a set of operators $\{\hat{\pi}_i\}$, called measurement operators, that satisfy
\begin{subequations}\label{POVM_Conditions}
\begin{alignat}{4}
    \hat{\pi}_i &\geq 0 ~~ \forall i,\\
    \sum \limits_i \hat{\pi}_i &= \hat{\mathbb{I}},
\end{alignat}
\end{subequations}
where $\hat{\mathbb{I}}$ denotes the identity operator. If a measurement $\{\hat{\pi}_i\}$ is performed on a state $\hat{\rho}_j$ and the measurement outcome is $k$ (considered a correct measurement outcome if $k=j$), then the state $\hat{\rho}_j$ is updated as follows \cite{kraus,barnett2009quantum}:
\begin{equation}\label{stateUpdate}
    \hat{\rho}_j \to \frac{\sqrt{\hat{\pi}_k} \hat{\rho}_j \sqrt{\hat{\pi}_k}^{\dagger}}{\text{Tr}(\hat{\pi}_k \hat{\rho}_j)}.
\end{equation}
Although this is not the unique form of allowed update rule for given $\hat{\pi}_k$, it is minimally disturbing and thus most appropriate for our purposes \cite{StateUpdate}. When a measurement $\{\hat{\pi}_i\}$ on a set of states $\{\hat{\rho}_i\}$ is performed, the probability of success is
\begin{equation}\label{pSuccGen}
    P_{\text{succ}} = \sum \limits_{i} p_i \text{Tr} \left( \hat{\pi}_i \hat{\rho}_i\right),
\end{equation}
where $p_i$ is the probability the input state is prepared in the state $\hat{\rho}_i$.

Finally, when considering multiple qubits, a basis that will turn out to be useful is the Schur basis \cite{SchurTransform}. The Schur basis states are denoted $|s,m_s\rangle |p_s\rangle$ where $p_s$ is what we call the ``path" degree of freedom. This basis is a consequence of Schur-Weyl duality which says \cite{SchurTransform}
\begin{equation}\label{2QubitSchurIsomorph}
    \!\left(\mathbb{C}^2\right)\!^{\otimes 2} \cong (\mathcal{Q}_{1} \otimes \mathcal{P}_{1}) \oplus (\mathcal{Q}_{0} \otimes \mathcal{P}_{0})
\end{equation}
for two qubits (i.e.\,\,two copies of the Hilbert space of a qubit) and
\begin{equation}\label{3QubitSchurIsomorph}
    \!\left(\mathbb{C}^2\right)\!^{\otimes 3} \cong (\mathcal{Q}_{\frac32} \otimes \mathcal{P}_{\frac32}) \oplus (\mathcal{Q}_{\frac12} \otimes \mathcal{P}_{\frac12})
\end{equation}
for three qubits. Here, $\mathcal{Q}_s ,\mathcal{P}_s \subset \left(\mathbb{C}^2\right)\!^{\otimes n}$ are the subspaces invariant under the action of the irreducible representations (irreps) of $\text{SU}(2)$ and $S_3$ respectively. These irreps of SU$(2),S_3$, and therefore subspaces $\mathcal{Q}_s,\mathcal{P}_s$ respectively, can be labelled by total spin $s$ since we are taking our qubits to be spin-$\frac12$ particles. With all this in mind, $\{|s,m_s\rangle \}$ is a basis for $\mathcal{Q}_s$ and $\{|p_s\rangle\}$ is a basis for $\mathcal{P}_s$. The reason we call $|p_s\rangle$ the path degree of freedom is that there exists a basis of $\mathcal{P}_s$ which corresponds to the different ways (or paths) by which a composite quantum system's state develops a spin-$s$ component via the spin 
addition of its constituent subsystems. This is the natural basis to work in for our problem because the states $|\varphi_k\rangle$ are completely unknown, so there is no preferred direction. This means that the states are maximally mixed within the subspaces $\mathcal{Q}_s$ corresponding to the irreps of $\text{SU}(2)$, and all the information is contained within the path degree of freedom: the $\mathcal{P}_s$ subspaces.

Explicitly, for two qubits, this basis relates to the computational basis as follows:
\begin{align}\label{2Schur}
    \begin{aligned}
        \mathcal{Q}_{1} \otimes \mathcal{P}_{1}: ~&\left|1,1\right\rangle = |00\rangle,\\
        & \left|1 , 0\right\rangle = \frac{1}{\sqrt{2}} (|01\rangle + |10\rangle ),\\
        &\left|1, -1\right\rangle = |11\rangle,\\
        \mathcal{Q}_{0} \otimes \mathcal{P}_{0}: ~& \left|0 , 0\right\rangle = \frac{1}{\sqrt{2}} (|01\rangle - |10\rangle )
    \end{aligned}
\end{align}
and for three qubits:
\begin{align}\label{3Schur}
    \begin{aligned}
        \mathcal{Q}_{\frac32} \otimes \mathcal{P}_{\frac32}: ~&\left|\frac32,\frac32\right\rangle = |000\rangle,\\
        &\left|\frac32, \frac12\right\rangle = \frac{1}{\sqrt{3}} (|100\rangle + |010\rangle + |001\rangle ),\\
        &\left|\frac32,-\frac12\right\rangle = \frac{1}{\sqrt{3}} (|011\rangle + |101\rangle + |011\rangle ),\\
        &\left|\frac32, -\frac32\right\rangle = |111\rangle,\\
        \mathcal{Q}_{\frac12} \otimes \mathcal{P}_{\frac12}: ~& \left|\frac12 , \frac12\right\rangle   |1\rangle = \frac{1}{\sqrt{6}} (|100\rangle + |010\rangle - 2|001\rangle ),\\
        &\left|\frac12 , -\frac12\right\rangle   |1\rangle = \frac{1}{\sqrt6} (-|011\rangle - |101\rangle + 2|110\rangle ),\\
        &\left|\frac12 , \frac12\right\rangle   |0\rangle =  \frac{1}{\sqrt2} (|100\rangle - |010\rangle),\\
        &\left|\frac12 , -\frac12\right\rangle   |0\rangle = \frac{1}{\sqrt2}(|101\rangle - |011\rangle).
    \end{aligned}
\end{align}
Here, we're using the abbreviation
\begin{equation}
    |i_1 i_2 \cdots i_n\rangle := |i_1\rangle \otimes |i_2\rangle\otimes \cdots \otimes |i_n\rangle
\end{equation}
for $i_k \in \{0,1\}$. Note also that when $\dim (\mathcal{P}_s) = 1$ we don't include $|p_s\rangle$. 
\vspace{-0.2cm}

\section{Optimal classification}

We begin by considering the optimal classification of two and three qubits separately. Part of the reason for doing this explicitly is to introduce some of the ideas and notation required for when we perform two sequential classifications. These results were derived previously in \cite{BARNETT2003,UnsupervisedClassification}.
\vspace{-0.5cm}

\subsection{Optimal classification of two qubits}\label{sec:twoOptimalqubits}

The aim of a classification of two unknown qubits is to determine whether these two qubits are the same as or different from one another \cite{BARNETT2003, UnsupervisedClassification}. That is, the aim is to distinguish between $|\varphi_0\rangle | \varphi_0\rangle,$ $|\varphi_0\rangle |\varphi_1\rangle,$ $|\varphi_1\rangle |\varphi_0\rangle$ and $|\varphi_1\rangle |\varphi_1\rangle$ (actually, we only distinguish between $|\varphi_0\rangle |\varphi_0\rangle$ and $|\varphi_0\rangle |\varphi_1\rangle$ as we will see). Let us begin by considering, mathematically, the form of the two-qubit states. Since $|\varphi_k\rangle$ are unknown qubits, to our knowledge, they are equally likely to be located at {\it any} point on the Bloch sphere. We therefore describe the two possible two-qubit states as mixed states using density operators as follows:
\begin{equation}
    \hat{\rho}_{ij} = \int |\varphi_i\rangle |\varphi_j\rangle\langle \varphi_i| \langle \varphi_j| \, d\varphi_0 d\varphi_1,
\end{equation}
where $i=0,~j\in\{0,1\}$, and the integral is with respect to the Haar measure and over the entire Bloch sphere. Note that we can always take $i=0$. This is because, when averaging over the Bloch sphere, all information about whether each qubit is $|\varphi_0\rangle$ or $|\varphi_1\rangle$ is lost and all that remains is information about their relative positions on the Bloch sphere. This means that $\hat{\rho}_{00} = \hat{\rho}_{11}$ and $\hat{\rho}_{01} = \hat{\rho}_{10}$.

The explicit form of these two states, in the Schur basis (Eq.\,(\ref{2Schur})), can be shown to be (see Appendix \ref{sec:Two-qubits} for more detail):
\begin{subequations}\label{2qubitStates}
    \begin{align}
    \hat{\rho}_{00} &= \frac13\left( |1,1\rangle\langle 1,1| + |1,0\rangle\langle 1,0| +|1,-1\rangle\langle 1,-1| \right),\\
    \hat{\rho}_{01} &= \frac14 \big( |1,1\rangle\langle 1,1| + |1,0\rangle\langle 1,0| + |1,-1\rangle\langle 1,-1| \nonumber\\ & \hspace{4.75cm} +|0,0\rangle\langle 0,0| \big).
    \end{align}
\end{subequations}
We can therefore observe that, here, a quantum classification of two unknown qubits corresponds to a quantum measurement (in general, a POVM) that distinguishes between the two states $\hat{\rho}_{00}, \hat{\rho}_{01}$. The optimal measurement to do this is made up of the projectors onto the totally symmetric and anti-symmetric subspaces invariant under $\text{SU}(2)$ respectively:
\begin{subequations}\label{Optimal2qubit}
\begin{align}
    \hat{P}_+ &= |1,1\rangle\langle 1,1| + |1,0\rangle\langle 1,0| +|1,-1\rangle\langle 1,-1|,\\
    \hat{P}_- &= |0,0\rangle\langle 0,0|,
\end{align}
\end{subequations}
where $\hat{P}_+ ~(\hat{P}_-)$ is the outcome associated with measuring the state $\hat{\rho}_{00} ~(\hat{\rho}_{01})$. Here we use the $+/-$ subscripts rather than $00/01$ with the hope that this makes the notation later in this paper less confusing to read. This measurement can be motivated by realising that $\hat{\rho}_{00}$ and $\hat{\rho}_{01}$ commute with one another, which means they have a common set of eigenstates. So we take the optimal measurement operators $\hat{P}_+,\hat{P}_-$ to be the (sum of) projectors onto the eigenstates with the largest eigenvalues of $\hat{\rho}_{00}, \hat{\rho}_{01}$ respectively. In other words, it is the Holevo-Helstrom measurement for
distinguishing between two quantum states \cite{nielsenAndChuang}. Using Eq.\,(\ref{pSuccGen}), the maximal probability of successfully classifying two equally-likely, unknown qubits is calculated as follows:
\begin{equation}
    P_{\text{succ}} = \frac12\!\left( \text{Tr}(\hat{P}_+ \hat{\rho}_{00}) + \text{Tr}(\hat{P}_- \hat{\rho}_{01}) \!\right)\!,
\end{equation}
where the $1/2$ comes from the two states $\hat{\rho}_{00}, \hat{\rho}_{01}$ being equiprobable. This results in a success rate of
\begin{equation}
    P_{\text{succ}} = \frac58 = 62.5\%.
\end{equation}
\vspace{-0.9cm}

\subsection{Optimal classification of three qubits}

Similarly to the two-qubit case, we begin by writing down the possible three-qubit states. In general, we once again express these states as
\begin{equation}
    \hat{\rho}_{ijk} = \int |\varphi_i\rangle |\varphi_j\rangle|\varphi_k\rangle \langle \varphi_i| \langle \varphi_j|\langle \varphi_k| \, d\varphi_0 d\varphi_1,
\end{equation}
where $i =0, ~j,k\in \{0,1\}$. As shown in Appendix \ref{sec:Three-qubit}, using the Schur basis (Eq.\,(\ref{3Schur})),
\begin{widetext}
\begin{subequations}\label{3QubitStates}
\begin{alignat}{4}
    \hat{\rho}_{000} &= \frac14\hat{\mathbb{I}}_{\frac32},\\
    \hat{\rho}_{001} &= \frac{1}{6}\hat{\mathbb{I}}_{\frac32}
    + \frac16 \hat{\mathbb{I}}_{\frac12} \otimes |1\rangle\langle 1|,\\
    \hat{\rho}_{010} &= \frac{1}{6} \hat{\mathbb{I}}_{\frac32} +
    \frac{1}{24}\hat{\mathbb{I}}_{\frac12}
    \otimes \left( |1\rangle - \sqrt{3}  |0\rangle \right)\!\! \left( \langle 1| - \sqrt{3} \langle 0| \right)\!,\\
    \hat{\rho}_{011} &= \frac{1}{6}\hat{\mathbb{I}}_{\frac32} +
    \frac{1}{24}\hat{\mathbb{I}}_{\frac12}
    \otimes\left( |1\rangle + \sqrt{3}  |0\rangle \right) \!\!\left( \langle 1| + \sqrt{3} \langle 0| \right)\!,
\end{alignat}
\end{subequations}
\end{widetext}
where $\hat{\mathbb{I}}_{s}$ is the identity operator on the subspace $\mathcal{Q}_{s}$. Note that $\hat{\mathbb{I}}_s, \hat{\mathbb{I}}_{s'}$ are orthogonal for $s \neq s'$. The optimal measurement that distinguishes the four states in Eq.\,(\ref{3QubitStates}) is
\begin{subequations}
\begin{alignat}{4}
    \hat{\pi}_{000} &= \hat{\mathbb{I}}_{\frac32},\\
    \hat{\pi}_{001} &= 
    \frac23 \hat{\mathbb{I}}_{\frac12}\otimes |1\rangle\langle 1|,\\
    \hat{\pi}_{010} &=
    \frac{1}{6}\hat{\mathbb{I}}_{\frac12}
    \otimes\left( |1\rangle - \sqrt{3}  |0\rangle \right)\!\! \left( \langle 1| - \sqrt{3} \langle 0| \right)\!,\\
    \hat{\pi}_{011} &=
    \frac{1}{6}\hat{\mathbb{I}}_{\frac12}
    \otimes\left( |1\rangle + \sqrt{3}  |0\rangle \right)\!\! \left( \langle 1| + \sqrt{3} \langle 0| \right)\!.
\end{alignat}
\end{subequations}
To motivate this, notice that $\hat{\rho}_{001},\hat{\rho}_{010},\hat{\rho}_{011}$ have $S_3$ permutation symmetry in their qubits. We can therefore require the optimal measurement to distinguish these three states to have this same symmetry. So, all we need to do is construct $\hat{\pi}_{000},\hat{\pi}_{001}$ to optimally distinguish between $\hat{\rho}_{000},\hat{\rho}_{001}$. From this, we can obtain $\hat{\pi}_{010}, \hat{\pi}_{011}$ via the $S_3$ symmetry mentioned. The construction of $\hat{\pi}_{000},\hat{\pi}_{001}$ follows the same reasoning as that of two-qubit measurement in Eq.\,(\ref{Optimal2qubit}) aside from the factor of $2/3$ in $\hat{\pi}_{001}$ which is required for completeness. Therefore, using this measurement and Eq.\,(\ref{pSuccGen}), the maximal probability of successfully distinguishing the (equally likely) states in Eq.\,(\ref{3QubitStates}) is
\begin{equation}
    P_{\text{succ}} = \frac{5}{12} \approx 41.7\%.
\end{equation}

\section{Measurement disturbance}\label{sec:seqOptimalMeas}

We thus consider what happens when we perform the optimal measurement (\ref{Optimal2qubit}) to classify two qubits, then add a third qubit and perform an optimal measurement on all three. In particular, we consider the case in which the outcome of the two-qubit measurement is known, and the measurement on all three is updated accordingly. After the first measurement has been performed with outcome $\hat{P}_j$, as discussed earlier with Eq.\,(\ref{stateUpdate}), the two qubit states update as follows:
\begin{equation}
    \hat{\rho}_{0i} \to \hat{\rho}_{0i}^j = \frac{\sqrt{\hat{P}_j} \hat{\rho}_{0i} \sqrt{\hat{P}_j}^{\dagger}}{\text{Tr} (\hat{P}_j \hat{\rho}_{0i})}.
\end{equation}
Following this, we add a third qubit, however, it is convenient to instead think of the situation as beginning with three qubits, and performing the measurement in Eq.\,(\ref{Optimal2qubit}) on the first two. With this in mind, following a measurement outcome of $\hat{P}_k$, the three qubit states are found using
\begin{equation}\label{rhoUpdate1}
    \hat{\rho}_{0ij}^k = \frac{\left(\!\sqrt{\hat{P}_k}\otimes \hat{\mathbbm{1}}\!\right)\! \hat{\rho}_{0ij} \!\left(\!\sqrt{\hat{P}_k}\otimes \hat{\mathbbm{1}}\!\right)^{\dagger}}{\text{Tr}(\hat{P}_k\otimes \hat{\mathbbm{1}} \hat{\rho}_{0ij})},
\end{equation}
where $\hat{\rho}_{0ij}$ are the states in Eq.\,(\ref{3QubitStates}) and $\hat{\mathbbm{1}}$ denotes the identity operator on a single qubit.

Explicitly, the states are as follows:
\begin{subequations}
\begin{alignat}{4}
    \quad \hat{\rho}^+_{000} &= \frac14\hat{\mathbb{I}}_{\frac32},\\
    \hat{\rho}^+_{001} &= \frac{1}{6}\hat{\mathbb{I}}_{\frac32}
    + \frac16 \hat{\mathbb{I}}_{\frac12}\otimes |1\rangle\langle 1|,\\
    \hat{\rho}_{01k}^+ &= \frac{2}{9}\hat{\mathbb{I}}_{\frac32} + \frac{1}{18} \hat{\mathbb{I}}_{\frac12}\otimes |1\rangle \langle 1|,\\
    \hat{\rho}_{000}^- &= \hat{0} = \hat{\rho}_{001}^-,\\
    \hat{\rho}_{01k}^- &= \frac12 \hat{\mathbb{I}}_{\frac12}\otimes |0\rangle \langle 0|
\end{alignat}
\end{subequations}
for $k = 0,1$. For each of the first measurement outcomes $\hat{P}_{\pm}$, we can therefore find the optimal measurement to be made up of the following projectors:
\begin{subequations}
\begin{alignat}{4}
    \hat{\pi}^+_{000} &= \hat{\mathbb{I}}_{\frac32},\\
    \hat{\pi}^+_{001} &= \hat{\mathbb{I}}_{\frac12 \otimes \frac12},\\
    \hat{\pi}_{01k}^+ &= \hat{0},\\
    \hat{\pi}^-_{000} &= \hat{\mathbb{I}}_{\frac32},\\
    \hat{\pi}^-_{001} &= \hat{0},\\
    \hat{\pi}_{01k}^- &= \frac12 \hat{\mathbb{I}}_{\frac12 \otimes \frac12},
\end{alignat}
\end{subequations}
where $\hat{\mathbb{I}}_{\frac12 \otimes \frac12}$ denotes the identity on the subspace $\mathcal{Q}_{\frac12} \otimes \mathcal{P}_{\frac12}$. These measurements can be motivated by the fact that $\hat{\pi}_i^{\pm}$ projects its corresponding state, $\hat{\rho}_i^{\pm}$, onto the components which are larger, or the same as, the same components in all the other states.

Now, the probability of a successful second measurement is given by
\begin{align}\label{Psucc2nd}
\begin{aligned}
    P_{\text{succ}}^{\text{2nd}} &= \sum \limits_{k \in \{+,-\}} \sum \limits_{i,j\in\{0,1\}} P(\hat{\rho}_{0ij})P(\hat{P}_k, \hat{\pi}_{0ij}^k|\hat{\rho}_{0ij}),\\
    &= \sum \limits_{k} \sum \limits_{i,j} P(\hat{\pi}_{0ij}^k | \hat{\rho}_{0ij}^k) P(\hat{P}_k|\hat{\rho}_{0ij}) P(\hat{\rho}_{0ij}) \!\\
    &= \frac14 \sum \limits_{k} \sum \limits_{i,j} \text{Tr}\!\left(\!\hat{\pi}_{0ij}^k \!\left(\!\sqrt{\hat{P}_k}\otimes \hat{\mathbbm{1}}\!\right)\! \hat{\rho}_{0ij} \!\left(\!\sqrt{\hat{P}_k}\otimes \hat{\mathbbm{1}}\!\right)\! \!\right)\!,
\end{aligned}
\end{align}
where $P(\hat{\rho}_{0ij})$ is the probability that the system is prepared in the state $\hat{\rho}_{0ij}$ (this is $1/4$ for all $i,j$), $P(\hat{P}_k, \hat{\pi}_{0ij}^k | \hat{\rho}_{0ij})$ denotes the probability that the first measurement outcome is $k$ and the second is $0ij$ given that the state was prepared in the state $\hat{\rho}_{0ij}$, and $P(\hat{\pi}_{0ij}^k | \hat{\rho}_{0ij}^k) \equiv P(\hat{\pi}_{0ij}^k | \hat{\rho}_{0ij}, \hat{P}_k)$. We therefore find that the probability of a successful second classification has been affected by an optimal first classification and has been reduced to the following value:
\begin{equation}
    P_{\text{succ}}^{\text{2nd}} = \frac{19}{48} \approx 39.6\%.
\end{equation}
Although this is a small reduction in the success rate of the three-qubit measurement, it demonstrates the principle of measurement disturbance caused by the intermediate classification.

\section{Weakening the intermediate measurement}

\subsection{Weak two-qubit measurement}

Our ultimate aim is to understand how a classification on two qubits affects our ability to perform a subsequent classification in general. So, instead of considering only the optimal measurement on two qubits, we interpolate between this and the weakest possible measurement: the identity measurement. This weakened measurement can be written as
\begin{align}\label{weakMeas1}
\begin{aligned}
    &\hat{\pi}_- = \alpha \hat{P}_- + \beta \hat{\mathbb{I}},\\
    &\hat{\pi}_+ = \alpha \hat{P}_+ + (1-\alpha-\beta) \hat{\mathbb{I}},\\
    &\text{such that } \alpha \in [0,1-\beta] \text{ and } \beta \in [0,1],
\end{aligned}
\end{align}
where the range of values $\alpha,\beta$ take come about due to the positivity condition of POVMs, given in Eq.\,(\ref{POVM_Conditions}a), as well as the convention we are adopting: we take the measurement outcome $\hat{\pi}_+$ ($\hat{\pi}_-$) to correspond to the measurement of the state $\hat{\rho}_{00}$ ($\hat{\rho}_{01}$). Note also that, by construction, this POVM is complete, as required (Eq.\,(\ref{POVM_Conditions}b)). To reduce future work, note that we can change between the two situations corresponding to different measurement outcomes by performing the swaps:
\begin{align}\label{swaps}
    \begin{aligned}
        \alpha &\to -\alpha,\\
        \beta &\to 1 - \beta.
    \end{aligned}
\end{align}
We conclude this subsection by noting that the probability of a successful two-qubit classification using the POVM in Eq.\,(\ref{weakMeas1}) is given by
\begin{equation}\label{Psucc1st}
    P_{\text{succ}}^{\text{1st}} = \frac12\!\left( 1 + \frac{\alpha}4 \right)\!,
\end{equation}where the superscript is included in anticipation of the second classification introduced in the next subsection.

\subsection{Adding a third qubit}

As before, after the first classification of two qubits has been performed, a third qubit, either $|\varphi_0\rangle$ or $|\varphi_1\rangle$, is added. In order to write down the resulting three-qubit state, just as in the case of an optimal intermediate measurement, it is convenient to think instead of the situation as an undisturbed three-qubit state $\hat{\rho}_{0ij}$ that is updated by the intermediate measurement on the first two qubits as
\begin{equation}\label{rhoUpdate}
    \hat{\rho}_{0ij}^{\pm} = \frac{(\sqrt{\hat{\pi}_{\pm}}\otimes \hat{\mathbbm{1}}) \hat{\rho}_{0ij} (\sqrt{\hat{\pi}_{\pm}}\otimes \hat{\mathbbm{1}})}{\text{Tr}(\hat{\pi}_{\pm}\otimes \hat{\mathbbm{1}} \hat{\rho}_{0ij})}.
\end{equation}

\noindent Explicitly, in the case when the measurement outcome on the first two qubits is $\hat{\pi}_-$, using similar techniques as those found in Eqs.\,(\ref{SchurUpdate}, \ref{Appendix001_2}) to find $\sqrt{\hat{\pi}_-}\otimes \hat{\mathbbm{1}}$ in the Schur basis, the states $\hat{\rho}_{0ij}^-$ can be shown to be
\begin{widetext}
\begin{subequations}
\begin{alignat}{4}
    \hat{\rho}^-_{000} &= \frac14\hat{\mathbb{I}}_{\frac32},\\
    \hat{\rho}^-_{001} &= \frac{1}{6}\hat{\mathbb{I}}_{\frac32}
    + \frac16 \hat{\mathbb{I}}_{\frac12}\otimes |1\rangle\langle 1|,\\
    \hat{\rho}^-_{010} &= \frac{4\beta}{6(\alpha + 4\beta)} \hat{\mathbb{I}}_{\frac32} + 
    \frac{1}{6(\alpha + 4\beta)}\hat{\mathbb{I}}_{\frac12}
    \otimes\left( \sqrt{\beta}   |1\rangle - \sqrt{3(\alpha + \beta)}  |0\rangle \right)\!\! \left( \sqrt{\beta}  \langle 1| - \sqrt{3(\alpha + \beta)} \langle 0| \right)\!,\\
    \hat{\rho}^-_{011} &= \frac{4\beta}{6(\alpha + 4\beta)}\hat{\mathbb{I}}_{\frac32} + 
    \frac{1}{6(\alpha + 4\beta)}\hat{\mathbb{I}}_{\frac12}
    \otimes\left( \sqrt{\beta}   |1\rangle + \sqrt{3(\alpha + \beta)}  |0\rangle \right)\!\! \left( \sqrt{\beta}  \langle 1| + \sqrt{3(\alpha + \beta)} \langle 0| \right)
\end{alignat}
\end{subequations}
\end{widetext}
with probabilities (derived in Appendix \ref{sec:DisturbedProbs})
\begin{subequations}
\begin{alignat}{4}
    p_{000}^-  &= p_{001}^- = \frac{2\beta}{\alpha+8\beta},\\
    p_{010}^- &= p_{011}^- = \frac{\alpha+4\beta}{2(\alpha+8\beta)}.
\end{alignat}
\end{subequations}
To find $\hat{\rho}^+_{0ij}$ we can just perform the swaps in Eq.\,(\ref{swaps}). 

In order to achieve our aim of classifying the resulting three-qubit system, we construct a measurement $\{ \hat{\pi}_i^- \}$ that distinguishes between the states $\{\hat{\rho}_i^-\}$ above (for additional detail, see Appendix \ref{sec:2ndMeasurementDeriv}). To do this, first, notice that the totally symmetric components ($s = 3/2$) of $\hat{\rho}_{001}^-, \hat{\rho}_{01l}^-$ are strictly less than that of $\hat{\rho}_{000}^-$. Further, $\hat{\rho}_{000}^-$ has no $s = \frac12$ components. This motivates the fact that the optimal way to distinguish $\hat{\rho}_{000}^-$ from the other states is to take
\begin{equation}
    \hat{\pi}_{000}^- = \hat{\mathbb{I}}_{\frac32}
\end{equation}
while keeping the remaining measurement operators in $(\mathcal{Q}_{\frac12} \otimes \mathcal{P}_{\frac12}) \otimes (\mathcal{Q}_{\frac12} \otimes \mathcal{P}_{\frac12})^*$, where $V^*$ denotes the dual space of $V$. Next, note that in the $s=\frac12$ subspaces, $\hat{\rho}_{001}, \hat{\rho}_{010}, \hat{\rho}_{011}$ have a mirror symmetric form in their path degree of freedom (spanned by $|p_{\frac12}\rangle = |0\rangle,|1\rangle$) as $p_{010}^- = p_{011}^-$ and the set is invariant under reflection about $|0\rangle$. Including $\hat{\pi}_{000}^-$ for completeness, the optimal measurement to distinguish these three states is known \cite{mirrorSym} and has the form
\begin{subequations}
\begin{alignat}{4}
    \hat{\pi}^{-}_{000} &= \hat{\mathbb{I}}_{\frac32},\\
    \hat{\pi}^{-}_{001} &= (1-a_{-}^2)\hat{\mathbb{I}}_{\frac12}\otimes |1\rangle\langle 1|,\\
    \hat{\pi}^{-}_{010} &= \frac12 \hat{\mathbb{I}}_{\frac12}
    \otimes ( a_{-}   |1\rangle - |0\rangle)\left( a_{-} \langle 1| -  \langle 0| \right),\\
    \hat{\pi}^{-}_{011} &= \frac12 \hat{\mathbb{I}}_{\frac12}
    \otimes ( a_{-}   |1\rangle + |0\rangle)\left( a_{-} \langle 1| +  \langle 0| \right),
\end{alignat}
\end{subequations}
where $a_{-} \in [0,1]$ to preserve positivity. A closed form analytic expression for $a_-$ in terms of the prior probabilities and overlaps of the states is given in \cite{mirrorSym}, which we use below. Once again, to obtain $\{\hat{\pi}_{i}^+\}$, we just perform the swaps in Eq.\,(\ref{swaps}). 

To utilise \cite{mirrorSym} we first must define a prior probability for the states $\hat{\rho}_{001}^-, \hat{\rho}_{01i}^-$ when projected into the $s=1/2$ subspace, and can then directly use the results of \cite{mirrorSym} to find the optimal value of the parameter $a_-$. Updating the prior probabilities gives
\begin{equation}
    p^- = \frac{3\alpha + 4\beta}{6(\alpha + 2\beta)},
\end{equation}
which is derived in Appendix \ref{sec:2ndMeasurementDeriv}. Using the analytical expression in \cite{mirrorSym} then gives (again more detail is given in Appendix \ref{sec:2ndMeasurementDeriv}):
\begin{equation}\label{aMinusCases2}
    a_- = \begin{cases}
    \sqrt{\frac{\alpha + \beta}{3\beta}} &\text{ if } \alpha \in [0,\text{min}\{ 1-\beta, 2\beta\}],\\
    \quad ~ ~ 1 &\text{ if } \alpha \in (2\beta,1-\beta] \text{ with } 2\beta < 1- \beta
    \end{cases}
\end{equation}
such that $\beta \in [0,1]$ as always. Note that the conditions $2\beta < 1 - \beta, ~\beta \in [0,1]$ can be rewritten as $\beta \in \left[0, \frac13\right)$. Similarly, when the outcome of the first measurement is $\hat{\pi}_+$, we arrive at 
\begin{equation}\label{aPlus}
    a_+ = \sqrt{\frac{1-\alpha - \beta}{3(1-\beta)}}
\end{equation}
for all valid $\alpha,\beta$. To achieve our aim and observe how the success probability of the first and second measurements compare to one another, we consider the two cases of Eq.\,(\ref{aMinusCases2}).

\vspace{-0.5cm}
\subsubsection{Case 1: $\alpha \in [0,\text{{\normalfont min}}\{ 1-\beta, 2\beta\}],\, \beta \in [0,1]$}\label{sec:case1}

Consider the first case in Eq.\,(\ref{aMinusCases2}), that is, when
\begin{align}
    \begin{aligned}
        a_- &= \sqrt{\frac{\alpha + \beta}{3\beta}},\\
        a_+ &= \sqrt{\frac{1-\alpha - \beta}{3(1-\beta)}}.
    \end{aligned}
\end{align}
Using Eq.\,(\ref{Psucc2nd}) with $\hat{P}_{\pm} \to \hat{\pi}_{\pm}$, in this region, it is straightforward, albeit requiring a little algebra, to show that the probability of a successful second classification stays constant at the optimal value for distinguishing three undisturbed qubits:
\begin{equation}\label{Psucc2ndConst}
    P_{\text{succ}}^{\text{2nd}} = \frac{5}{12}
\end{equation}
for all $\alpha \in [0,\text{min}\{ 1-\beta, 2\beta\}], \,\beta \in [0,1]$.

\vspace{-0.5cm}
\subsubsection{Case 2: $\alpha \in (2\beta, 1 - \beta], \beta \in \left[0,\frac13\right)$}\label{sec:case2}

Considering now the second case in Eq.\,(\ref{aMinusCases2}), let $a_- = 1$ and $a_+$ be as written in Eq.\,(\ref{aPlus}). Once again, using Eq.\,(\ref{Psucc2nd}) with $\hat{P}_{\pm} \to \hat{\pi}_{\pm}$, after a little algebra, we find
\begin{equation}\label{Psucc2ndAlphaBeta}
    P_{\text{succ}}^{\text{2nd}} = \frac{5}{12} - \frac{\beta}{12} - \frac{\alpha}{48} + \frac1{24} \sqrt{3\beta(\alpha + \beta)}.
\end{equation}
Now, we want $P_{\text{succ}}^{\text{2nd}}$ to be its optimal value for each value of $P_{\text{succ}}^{\text{1st}}$. Since $P_{\text{succ}}^{\text{1st}}$ has the form given in Eq.\,(\ref{Psucc1st}) (linear in $\alpha$ alone), to do this, we hold $\alpha$ constant, and maximise $P_{\text{succ}}^{\text{2nd}}$ with respect to $\beta$. This occurs when
\begin{equation}
    \beta = -\frac{3\alpha}{2}
\quad \text{or} \quad \beta = \frac{\alpha}{2}.
\end{equation}

The first option only holds when $\alpha = 0 \not \in (2\beta, 1-\beta]$. The second option corresponds to the boundary of the two scenarios in Eq.\,(\ref{aMinusCases2}) - that is, when $\alpha = 2\beta$. This tells us that for $\alpha > 2\beta$, there are no stationary points with respect to $\beta$, and we must therefore look to the boundaries of $\beta$: $\beta = 0$ or $\beta = 1 - \alpha$. However, the optimal boundary can be shown to be $\beta = 1 - \alpha$ when we notice that $P_{\text{succ}}^{\text{2nd}}$ is monotonically increasing with respect to $\beta$ in the region $\alpha \in (2\beta, 1-\beta]$, $\beta \in \left[ 0 ,\frac13\right]$. This can be shown using the fact that there are no stationary points in this region, so it must therefore be monotonically increasing or decreasing, along with the fact there exists a point (e.g $(\alpha,\beta) = (5/6, 1/6)$) in this region such that $\frac{\partial P_{\text{succ}}^{\text{2nd}}}{\partial \beta} > 0$. So, using $\beta = 1-\alpha$ along with Eq.\,(\ref{Psucc1st}), we find the optimal probability of success in this region to be
\begin{equation}\label{Psucc2ndCandidate2}
    P_{\text{succ}}^{\text{2nd}} = \frac{1}{12}+\frac{P_{\text{succ}}^{\text{1st}}}{2} +  \frac{1}{24}\sqrt{3(5-8P_{\text{succ}}^{\text{1st}})}.
\end{equation}

We can re-express the boundaries in $P_{\text{succ}}^{\text{2nd}}$ in terms of $P_{\text{succ}}^{\text{1st}}$ by noting that we'd like Eq.\,(\ref{Psucc2ndConst}) to be the success rate for as large a region as possible. This can be seen by noting that Eq.\,(\ref{Psucc2ndAlphaBeta}) can be rewritten as
\begin{equation}
    P_{\text{succ}}^{\text{2nd}} = \frac{5}{12} - \frac{1}{48}\! \left(\!\sqrt{\alpha + \beta} - \sqrt{3\beta}\right)^2
\end{equation}
and therefore is less than or equal to the optimal value of $5/12$. So, to make the region in which Eq.\,(\ref{Psucc2ndConst}) is true as large as possible, we must maximise $\text{min}\{2\beta, 1- \beta\}$. That is, when $\beta = 1/3$ and so $\alpha \in \left[ 0, \frac13\right]$. Therefore, using Eq.\,(\ref{Psucc1st}), we take $P_{\text{succ}}^{\text{2nd}}$ to be given by Eq.\,(\ref{Psucc2ndConst}) when $P_{\text{succ}}^{\text{1st}} \in \left[0,\frac{7}{12}\right]$, and by Eq.\,(\ref{Psucc2ndCandidate2}) when $P_{\text{succ}}^{\text{1st}} \in \left(\frac{7}{12}, \frac58 \right]$.

\begin{figure*}
    \centering
    \includegraphics[width=\textwidth]{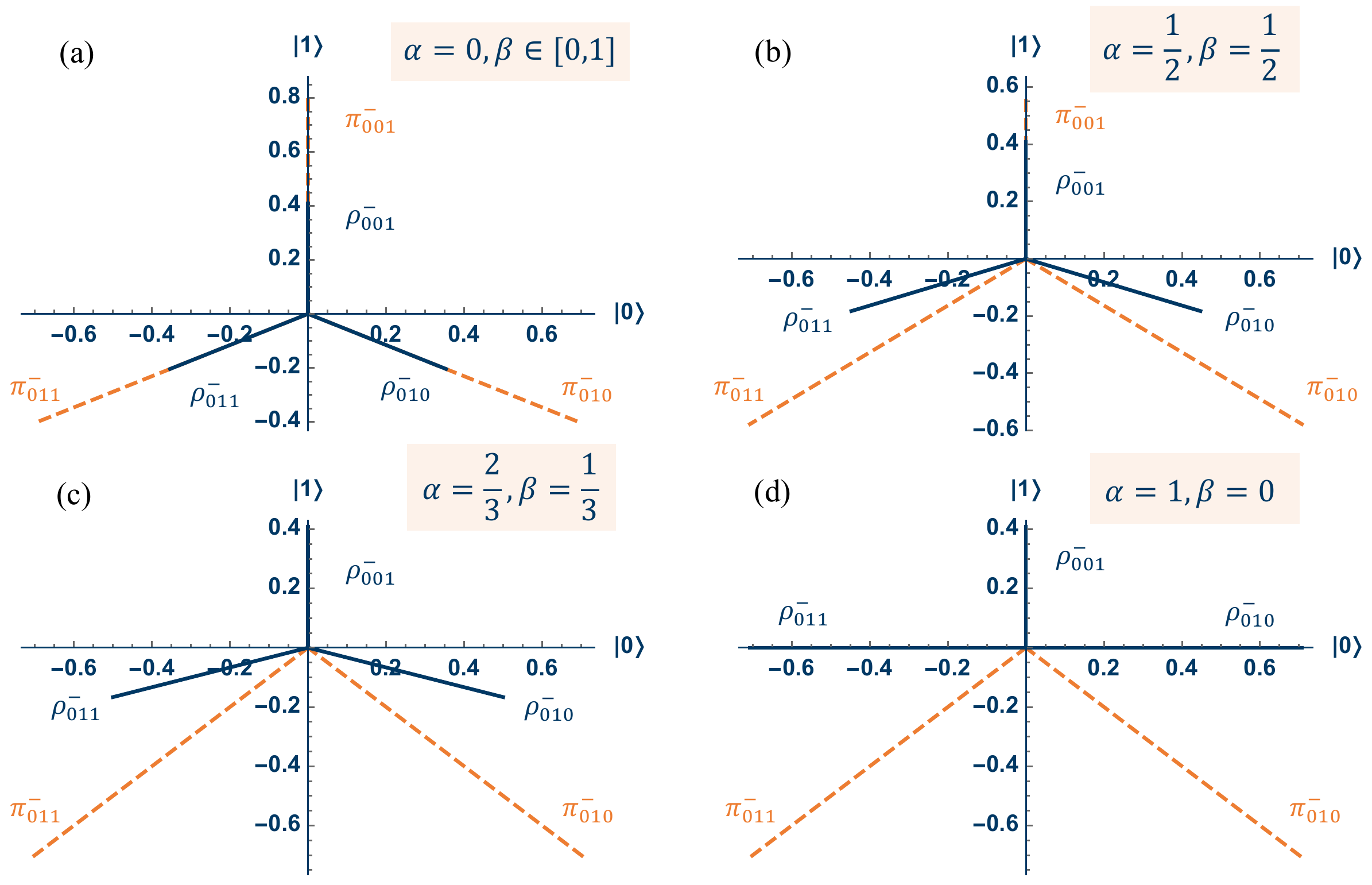}
    \caption{Plots showing the effect that a measurement on the first two qubits of a system, with outcome $\hat{\pi}_-$ (for various values of $\alpha,\beta$), has on the three-qubit states and measurement operators $\{ \hat{\rho}^-_{001}, \hat{\rho}^-_{010}, \hat{\rho}^-_{011}\}$ and $\{ \hat{\pi}^-_{001}, \hat{\pi}^-_{010}, \hat{\pi}^-_{011}\}$ respectively. Note that the states and measurement operators vary with the strength of the 2-qubit measurement until $\alpha = \frac23,~ \beta = \frac13$ (FIG.\,1(c)), after which, while the states continue to change, the measurement operators stay constant. FIG.\,1(c) corresponds to the boundary between the constant and non-constant regions in FIG.\,\ref{fig:P2nd_Vs_P1st_tradeoff}. Further, FIG.\,1(a) corresponds to the case in which no measurement is performed on the first two qubits and FIG.\,1(d) to when the optimal measurement is performed on the first two qubits.} 
    \label{fig:States&Measurements}
\end{figure*}

To gain some intuition as to how the three qubit states and second measurement vary with the strength of the first measurement, we can plot their $s= \frac12$ path components - that is, their components when restricted to the subspace $\mathcal{P}_{\frac12}$. Further, since $\hat{\rho}_{000}^{\pm}, \hat{\pi}_{000}^{\pm}$ are left invariant by the first measurement, no information is gained by considering them, so we only need look at the remaining states and measurement operators. FIG.\,\ref{fig:States&Measurements} shows how the states $\hat{\rho}_{001}^-,\hat{\rho}_{010}^-,\hat{\rho}_{011}^-$ and measurement operators $\hat{\pi}_{001}^-,\hat{\pi}_{010}^-,\hat{\pi}_{011}^-$ compare to one another for various values of $\alpha, \beta$. Note the mirror symmetry of the states and measurement operators in their $|0\rangle$ components as discussed earlier when constructing the optimal measurement of the three-qubit states. Further, the adjustment of the second measurement compensates for the disturbance caused by the first measurement in the region $\alpha \in [0,\text{min}\{ 1-\beta, 2\beta\}], \beta \in [0,1]$.
\begin{figure*}
    \centering
    \includegraphics[width=\textwidth]{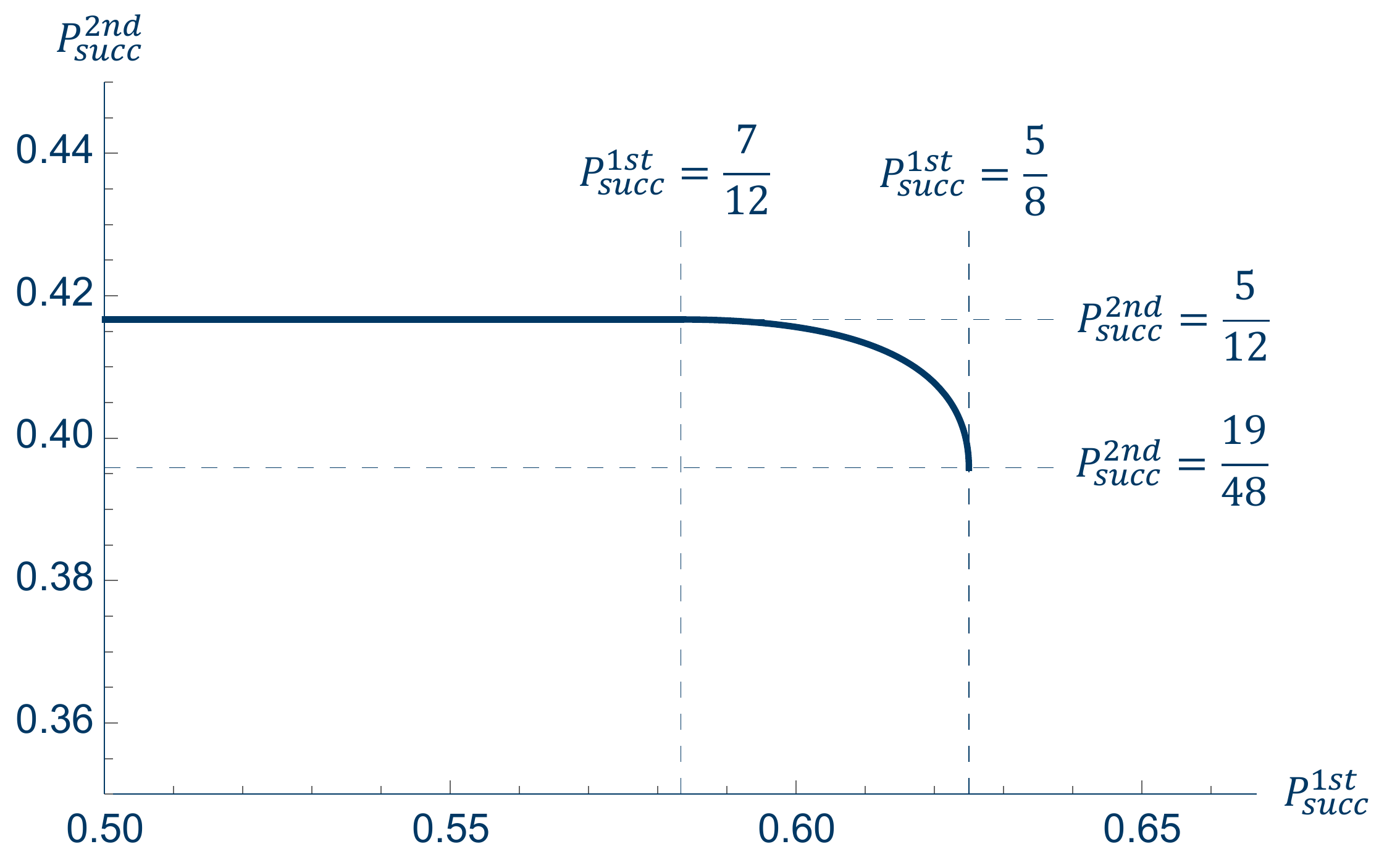}
    \caption{Plot of the tradeoff between the success rates of the first binary classification of two qubits and the second binary classification when a third qubit is added. The probability of success of the first (second) measurement is denoted $P_{\text{succ}}^{\text{1st}}$ $(P_{\text{succ}}^{\text{2nd}})$.}
    \label{fig:P2nd_Vs_P1st_tradeoff}
\end{figure*}

\vspace{-0.02cm}
\section{Results}

Summarising what we have found, the tradeoff between the first and second classification is given by
\begin{widetext}
\begin{equation}\label{tradeoff}
    P_{\text{succ}}^{\text{2nd}} = \begin{cases}
    \quad \quad \quad \quad \quad \quad \frac{5}{12} &\text{ if } P_{\text{succ}}^{\text{1st}} \in \left[0,\frac{7}{12} \right],\\
    \frac{1}{12}+\frac{P_{\text{succ}}^{\text{1st}}}{2} +  \frac{1}{24}\sqrt{3(5-8P_{\text{succ}}^{\text{1st}})} &\text{ if } P_{\text{succ}}^{\text{1st}} \in \left(\frac{7}{12},\frac58 \right].
    \end{cases}
\end{equation}
\end{widetext}
\vspace{-1cm}
A plot of this tradeoff can be seen in FIG.\,\ref{fig:P2nd_Vs_P1st_tradeoff}. Let's note some points of interest. Firstly, when we require the second measurement to be optimal, the best first measurement occurs when $\alpha = \frac23$ and $\beta = 1-\alpha = \frac13$. Here,
\begin{subequations}
\begin{alignat}{4}
    P_{\text{succ}}^{\text{1st}} &= \frac{7}{12} \approx 58.3\%,\\
    P_{\text{succ}}^{\text{2nd}} &= \frac{5}{12} \approx 41.7\%.
\end{alignat}
\end{subequations}
So the success rate of the first measurement, under the requirement that $P_{\text{succ}}^{\text{2nd}}$ is optimal, ranges from $\frac12$ to $\frac{7}{12}$. It is worth reiterating that the transition from optimal to sub-optimal second-measurement success rate occurs at the boundary of the two cases in Eq.\,(\ref{tradeoff}) or Eq.\,(\ref{aMinusCases2}). That is, given a first outcome of $\hat{\pi}_-$, when the second measurement stops varying with respect to $\alpha, \beta$ as can be seen in FIG.\,\ref{fig:States&Measurements}. 

The next point to consider is when we optimise $P_{\text{succ}}^{\text{1st}}$. Here $\alpha = 1$ and $\beta = 1-\alpha = 0$ which means that
\begin{subequations}
\begin{alignat}{4}
    P_{\text{succ}}^{\text{1st}} &= \frac58 = 62.5\%,\\
    P_{\text{succ}}^{\text{2nd}} &= \frac{19}{48} \approx 39.6\%,
\end{alignat}
\end{subequations}
as was found in Sections \ref{sec:twoOptimalqubits} and \ref{sec:seqOptimalMeas}.
This limited success rate of the second measurement can perhaps be expected due to the fact the $\rho_{01k}^-$ are parallel to one another as can be seen in FIG.\,\ref{fig:States&Measurements}(d).

\section{Conclusion}

To summarise, we considered a base case in the tradeoff between two sequential unsupervised quantum learning tasks. In particular, we looked at the situation in which there were initially two qubits that could each be in one of two unknown quantum states. Once a binary classification of varying success rate, corresponding to a quantum measurement of varying strength, had been performed, a third qubit was added and the optimal classification on all three qubits was then performed. We found that, although a binary classification of two unknown qubits causes measurement disturbance which can degrade the performance of an optimal classifier on all three qubits, there is a large regime in which the performance remains unaffected. In this regime, the final measurement may be adjusted to fully mitigate the disturbance caused by the first measurement. That is, the success rate of the first classification can range from that of a guess, $P_{\text{succ}}^{\text{1st}} = 1/2$ to $P_{\text{succ}}^{\text{1st}} = 7/12$ without causing the success rate of the second classification to deviate from its optimal value of $P_{\text{succ}}^{\text{2nd}} = 5/12$. When $P_{\text{succ}}^{\text{1st}}$ is further improved, however, $P_{\text{succ}}^{\text{2nd}}$ decreases non-linearly to a success rate of $19/48$ as $P_{\text{succ}}^{\text{1st}}$ increases to its optimal value of $5/8$.

This work provides an indication that sequential unsupervised classifications of quantum data can be performed. Further, depending on the strength of an earlier classification, a later classification's ability need not be compromised. Having said this, this work also highlights that there are non-trivial tradeoffs between sequential unsupervised quantum learning tasks which, although small in this base case, may be more considerable in more complicated scenarios. Here we have considered the simplest possible example of a quantum learning task in which a measurement disturbance tradeoff exists between performance on a subset of the data provided and performance on the whole dataset. We have fully characterized this tradeoff. This is a peculiarly quantum effect due to fundamental features of quantum mechanics, which is not present in classical machine learning. 

This is just the first step in exploring this tradeoff in learning tasks, and more work is required to fully understand the limitations imposed by quantum mechanics on sequential learning. For example the next natural step would be to consider starting with $n$ unknown qubits of two types and, following a classification of them, adding 1 or more extra qubits to be subsequently classified. Further, one could look at the case in which a larger number of options of qubit (or $d$-dimensional qudit) to choose between. Another path to take could be the supervised analogue of the content of this paper, with labelled qubits being given as a training set used to classify future ones. In addition, there are a range of learning scenarios, including partially or fully supervised learning, and reinforcement learning, in which similar effects may be explored. We leave these considerations for future work.

\begin{acknowledgements}
The authors acknowledge The Engineering and Physical Sciences Research Council and the UK National Quantum Technologies Programme via the QuantIC Quantum Imaging Hub (EP/T00097X/1). Sarah Croke is supported by a Leverhulme Fellowship (RF-2020-397).
\end{acknowledgements}

\appendix

\section{Derivation of undisturbed states}

Let us first derive the two-qubit states in Eq.\,(\ref{2qubitStates}). To do this, we use some representation theory. For our purposes, we define the representation $(\mathbf{Q}_n, \mathcal{Q})$ of $SU(2)$ such that for any $U \in SU(2)$,
\begin{equation}
    \mathbf{Q}_n(U) |i_1\rangle \cdots |i_n\rangle := U^{\otimes n} |i_1\rangle \cdots |i_n\rangle.
\end{equation}
We also require the following representation of the symmetric group $S_n$:
\begin{equation}
    \mathbf{P}(\sigma) |i_1\rangle \cdots |i_n\rangle := |i_{\sigma^{-1}(1)}\rangle \cdots |i_{\sigma^{-1}(n)}\rangle
\end{equation}
such that $\sigma \in S_n$. For instance, for $n=3$, $\sigma = (123)$,
\begin{equation}
    \mathbf{P}((123))|i_1 i_2 i_3 \rangle = |i_3 i_1 i_2\rangle.
\end{equation}

\subsection{Two-qubit states}\label{sec:Two-qubits}

Beginning with $\hat{\rho}_{00}$, using the $SU(2)$ invariance of the Haar measure, notice that
\begin{align}
\begin{aligned}
    \mathbf{Q}_2(U)\hat{\rho}_{00}\mathbf{Q}_2^{\dagger}(U) &= \mathbf{Q}_2(U) \left(\int |\varphi_0 \varphi_0 \rangle \langle \varphi_0 \varphi_0| d\varphi_0 \right)\mathbf{Q}_2^{\dagger}(U)\\
    &= \int |\varphi_0 \varphi_0 \rangle \langle \varphi_0 \varphi_0| d\varphi_0 = \hat{\rho}_{00}.
\end{aligned}
\end{align}
So, by Schur's Lemma, there exist bases, for example the Schur basis given in Eq.\,(\ref{2Schur}), such that
\begin{equation}
    \hat{\rho}_{00} = \alpha_1 \hat{\mathbb{I}}_1 \oplus \alpha_0 \hat{\mathbb{I}}_0.
\end{equation}
Recalling Eq.\,(\ref{2QubitSchurIsomorph}), note that the subscripts reference the total spin of the subspaces.

However, notice that for any qubit $|\varphi_0\rangle = a|0\rangle + b|1\rangle$,
\begin{equation}
    |\varphi_0 \varphi_0 \rangle = a^2 |00\rangle + ab\left( |01\rangle + |10\rangle \right) + b^2 |11\rangle.
\end{equation}
So, comparing with Eq.\,(\ref{2Schur}), $|\varphi_0 \varphi_0 \rangle$ lives entirely in $\mathcal{Q}_1 \otimes \mathcal{P}_1$. This implies that $\alpha_0 = 0$ and therefore, by the normalisation of $\hat{\rho}_{00}$,
\begin{align}
\begin{aligned}
    \hat{\rho}_{00} &= \frac13 \hat{\mathbb{I}}_{1}\\
    &= \frac13\left( |1,1\rangle\langle 1,1| + |1,0\rangle\langle 1,0| +|1,-1\rangle\langle 1,-1| \right).
\end{aligned}
\end{align}

Next, for $\hat{\rho}_{01}$, notice that with a slight abuse of notation,
\begin{align}
    \begin{aligned}
        \hat{\rho}_{01} &= \int |\varphi_0 \varphi_1\rangle \langle \varphi_0 \varphi_1| d\varphi_0 d\varphi_1\\
        &= \int |\varphi_0\rangle\langle \varphi_0| \otimes |\varphi_1\rangle\langle \varphi_1| d\varphi_0 d\varphi_1 \\
        &= \int |\varphi_0\rangle\langle \varphi_0|d\varphi_0 \otimes \int |\varphi_1\rangle\langle \varphi_1|  d\varphi_1\\
        &= \frac14 \hat{\mathbbm{1}} \otimes \hat{\mathbbm{1}}
    \end{aligned}
\end{align}
where the last equality is obtained by the invariance of $\int |\varphi_i\rangle\langle \varphi_i|d\varphi_i$ under $\mathbf{Q}_1(SU(2))$ (or, more physically, due to each of the integrals describing a maximally mixed qubit) and the $1/4$ is required for normalisation. Therefore, $\hat{\rho}_{01}$ is proportional to the identity on $\left(\mathbb{C}^2\right)\!^{\otimes 2}$ and hence, we can rewrite it as the identity in the Schur basis:
\begin{align}
    \begin{aligned}
        \hat{\rho}_{01} &= \frac14 \hat{\mathbb{I}}_1 \oplus \hat{\mathbb{I}}_0\\
        &= \frac14 \big( |1,1\rangle\langle 1,1| + |1,0\rangle\langle 1,0| + |1,-1\rangle\langle 1,-1| \nonumber\\ & \hspace{4.75cm} +|0,0\rangle\langle 0,0| \big).
    \end{aligned}
\end{align}

\subsection{Three-qubit states}\label{sec:Three-qubit}

Similar arguments to the two-qubit case tell us that we can write the following:
\begin{equation}
    \hat{\rho}_{000} = \alpha_{\frac32} \hat{\mathbb{I}}_{\frac32} \oplus \alpha_{\frac12} \hat{\mathbb{I}}_{\frac12} \oplus \alpha_{\frac12}' \hat{\mathbb{I}}_{\frac12}
\end{equation}
due to its commutivity with all the elements of $\mathbf{Q}_3(SU(2))$. To motivate the presence of two copies of the $\mathcal{Q}_{\frac12}$ space implied here, recall that the addition of three spin-half particles results in a system with {\it two} orthogonal spin-half components. To find $\alpha_{\frac12}, \alpha_{\frac12}'$, notice that for all $\sigma \in S_3$,
\begin{equation}
    \mathbf{P}(\sigma) \hat{\rho}_{000} \mathbf{P}^{\dagger}(\sigma) = \hat{\rho}_{000}.
\end{equation}
This implies $\hat{\rho}_{000}$ lives entirely within $(\mathcal{Q}_{\frac32} \otimes \mathcal{P}_{\frac32})\otimes (\mathcal{Q}_{\frac32} \otimes \mathcal{P}_{\frac32})^*$, for if it didn't, it would have a component within $(\mathcal{Q}_{\frac12} \otimes \mathcal{P}_{\frac12})\otimes (\mathcal{Q}_{\frac12} \otimes \mathcal{P}_{\frac12})^*$ and would therefore not be acted on trivially by $\mathbf{P}(S_3)$ since the irrep that $\mathcal{P}_{\frac12}$ is invariant under is not trivial. It therefore follows that $\alpha_{\frac12},\alpha_{\frac12}' = 0$ and
\begin{equation}
    \hat{\rho}_{000} = \frac14 \hat{\mathbb{I}}_{\frac32},
\end{equation}
where, again, $1/4$ is the normalisation constant.

Now, for $\hat{\rho}_{001}$, using a similar technique to $\hat{\rho}_{01}$,
\begin{align}
    \begin{aligned}\label{Appendix001}
    \hat{\rho}_{001} =& \int |\varphi_0 \varphi_0 \varphi_1\rangle \langle \varphi_0 \varphi_0 \varphi_1| d\varphi_0 d\varphi_1\\
    =& \int |\varphi_0 \varphi_0\rangle \langle \varphi_0 \varphi_0| d\varphi_0 \otimes \int  |\varphi_1\rangle \langle  \varphi_1| d\varphi_1\\
    =& \frac12 \hat{\rho}_{00} \otimes \hat{\mathbbm{1}}\\
    =& \frac16 \left(|1,1\rangle\langle 1,1| + |1,0\rangle\langle 1,0| + |1,-1\rangle\langle 1,-1|\right)\\
    &~~\otimes \left( \left| \frac12, \frac12 \right\rangle \left\langle \frac12, \frac12 \right| +\left| \frac12, -\frac12 \right\rangle \left\langle \frac12, -\frac12 \right| \right),
    \end{aligned}
\end{align}
where the prefactors are determined using similar ideas to before. In order to rewrite this in the Schur basis of $(\mathcal{Q}_{\frac32} \otimes \mathcal{P}_{\frac32})\otimes (\mathcal{Q}_{\frac12} \otimes \mathcal{P}_{\frac12})$, we use the following \cite{SchurTransform}:
\begin{multline}\label{SchurUpdate}
    |s,m\rangle|p\rangle \otimes \left| \frac12, \pm\frac12 \right\rangle \\
    \to \sqrt{\frac{s\pm m + 1}{2s + 1}} \left| s+\frac12, m\pm \frac12 \right\rangle|p,0\rangle \\
    \mp \sqrt{\frac{s \mp m}{2s + 1}} \left| s - \frac12, m \pm \frac12 \right\rangle |p,1\rangle.
\end{multline}
For our case, $s = 1, m \in \{ 1,0,-1\}$ and $p$ has been omitted since $\dim \mathcal{P}_1 = 1 =  \dim \mathcal{P}_0$. Applying Eq.\,(\ref{SchurUpdate}) to Eq.\,(\ref{Appendix001}), we obtain
\begin{equation}\label{Appendix001_2}
    \hat{\rho}_{001} = \frac16\left( \hat{\mathbb{I}}_{\frac32} + \hat{\mathbb{I}}_{\frac12} \otimes |1\rangle \langle 1 | \right),
\end{equation}
where $\alpha = 1/6$ was found by again requiring $\text{Tr}(\hat{\rho}_{001}) = 1$.

Finally, we can find $\hat{\rho}_{010},\hat{\rho}_{011}$. Noting that,
\begin{align}
\begin{aligned}
    \hat{\rho}_{011} &= \int |\varphi_0 \varphi_1 \varphi_1\rangle \langle \varphi_0 \varphi_1 \varphi_1| d\varphi_0 d\varphi_1\\
    &= \int |\varphi_1 \varphi_0 \varphi_0\rangle \langle \varphi_1 \varphi_0 \varphi_0| d\varphi_0 d\varphi_1 = \hat{\rho}_{100},
\end{aligned}
\end{align}
to obtain $\hat{\rho}_{010}, \hat{\rho}_{011}$, notice that we just have to permute the qubits in $\hat{\rho}_{001}$. To do this, first notice that $\hat{\mathbb{I}}_{\frac32}$ is invariant under permutations of qubits since $\mathbf{P}(S_3)$ acts trivially on $\mathcal{P}_{\frac32}$. Therefore, the only part of $\hat{\rho}_{001}$ affected by permutations is $\hat{\mathbb{I}}_{\frac12} \otimes |1\rangle\langle 1|$.

Intuitively, we can guess the form of $\hat{\rho}_{010}, \hat{\rho}_{011}$ by the fact that,
\begin{equation}
    \cdots \hat{\rho}_{001} \xrightarrow[]{(123)} \hat{\rho}_{010} \xrightarrow[]{(123)} \hat{\rho}_{011} \xrightarrow[]{(123)} \hat{\rho}_{001} \cdots,
\end{equation}
where $(123)\in S_3$ is a 3-cycle. Since permutations only have an affect on the path components of the states, it seems that $\hat{\rho}_{001}, \hat{\rho}_{010}, \hat{\rho}_{011}$ should be evenly distributed in the 2-dimensional space $\mathcal{P}_{\frac12}$. That is, since each state can be accessed by repeated application of the permutation $(123)$, we'd expect each state be accessible by the repeated application of some 2D-transformation (on $\mathcal{P}_{\frac12}$). In particular, since $\hat{\rho}_{001}$ is known, we might guess that the remaining states could be found by rotating its $\mathcal{P}_{\frac12}$ component by $2\pi/3$. This indeed results in the states given in Eq.\,(\ref{3QubitStates}).

More explicitly, we can derive $\hat{\rho}_{010},\hat{\rho}_{011}$ using the following three steps:
\begin{enumerate}
    \item Rewrite $\hat{\rho}_{001}$ in the computational basis (using Eq.\,(\ref{3Schur})).
    \item Permute the qubits $|i_1 i_2 i_3 \rangle \to |i_2 i_3 i_1 \rangle$ to obtain $\hat{\rho}_{010}$ in the computational basis.
    \item Rewrite the state in the Schur basis given in Eq.\,(\ref{3Schur}).
\end{enumerate}

\section{Derivation of updated states and measurements}

\subsection{Updated prior probabilities}\label{sec:DisturbedProbs}

Assuming the outcome of the measurement on the first two qubits is $-$, the probabilities of the disturbed states $\hat{\rho}_{000}^-, \hat{\rho}_{001}^-, \hat{\rho}_{010}^-, \hat{\rho}_{011}^-$ occurring are given by $p_{000}^-,p_{001}^-,p_{010}^-,p_{011}^-$ respectively, such that
\begin{equation}
    p_{0ij}^- := P(\hat{\rho}_{0ij} | \hat{\pi}_- \otimes \mathbbm{1}).
\end{equation}
Using Bayes' theorem, this can be written as
\begin{align}
    p_{0ij}^- &= \frac{P( \hat{\pi}_- \otimes \mathbbm{1}| \hat{\rho}_{0ij})P(\hat{\rho}_{0ij})}{P( \hat{\pi}_- \otimes \mathbbm{1})} \nonumber\\
    &= \frac{P( \hat{\pi}_- | \hat{\rho}_{0i})}{4P( \hat{\pi}_-)}, 
\end{align}
where the second equality is obtained using the fact that $P(\hat{\rho}_{0ij}) = 1/4$, and that the third qubit is acted on only by the identity and therefore does not change any of the probabilities.

So, by noting that
\begin{align}
\begin{aligned}
    P(\hat{\pi}_- | \hat{\rho}_{00}) &= \text{Tr} ( \hat{\pi}_-  \hat{\rho}_{00} ) = \beta,\\
    P(\hat{\pi}_- | \hat{\rho}_{01}) &= \text{Tr} ( \hat{\pi}_-  \hat{\rho}_{01} ) = \frac14 (\alpha + 4\beta),
\end{aligned}
\end{align}
and therefore,
\begin{align}
    P( \hat{\pi}_-) &= P(\hat{\pi}_- | \hat{\rho}_{00})P(\hat{\rho}_{00})  + P(\hat{\pi}_- | \hat{\rho}_{01})P(\hat{\rho}_{01}) \nonumber \\
    &= \frac18 (\alpha + 8\beta),
\end{align}
we find that
\begin{subequations}
\begin{alignat}{4}
    p_{000}^-  &= p_{001}^- = \frac{2\beta}{\alpha+8\beta},\\
    p_{010}^- &= p_{011}^- = \frac{\alpha+4\beta}{2(\alpha+8\beta)}.
\end{alignat}
\end{subequations}

\subsection{Second measurement}\label{sec:2ndMeasurementDeriv}

Again, assuming the outcome of the first classification was $-$, recall that distinguishing $\hat{\rho}_{000}^-$ from the other states is done optimally by letting $\hat{\pi}_{000}^-$ be the projector onto the $s=3/2$ space. This leads to the measurement that best distinguishes $\hat{\rho}_{001}^-,\hat{\rho}_{010}^-,\hat{\rho}_{011}^-$ being entirely contained in the $s=1/2$ space $(\mathcal{Q}_{\frac12} \otimes \mathcal{P}_{\frac12}) \otimes (\mathcal{Q}_{\frac12} \otimes \mathcal{P}_{\frac12})^*$. Further, since the $\mathcal{Q}_{\frac12} \otimes \mathcal{Q}_{\frac12}^*$ component of each of $\hat{\rho}_{001}^-,\hat{\rho}_{010}^-,\hat{\rho}_{011}^-$ is the identity, all the information about how they differ is contained in $\mathcal{P}_{\frac12} \otimes \mathcal{P}_{\frac12}^*$. So we can rephrase this as a state discrimination problem of the following states:
\begin{subequations}
\begin{alignat}{4}
    |\psi_{001}^-\rangle &= \frac{\mathcal{N}_{001}^-}{\sqrt6} |1\rangle,\\
    |\psi_{010}^-\rangle &= \frac{\mathcal{N}_{010}^-}{\sqrt{6(\alpha+4\beta)}}\left(\sqrt{\beta} |1\rangle - \sqrt{3(\alpha + \beta)}|0\rangle \right),\\
    |\psi_{011}^-\rangle &= \frac{\mathcal{N}_{011}^-}{\sqrt{6(\alpha+4\beta)}}\left(\sqrt{\beta} |1\rangle + \sqrt{3(\alpha + \beta)}|0\rangle \right),
\end{alignat}
\end{subequations}
where $\mathcal{N}_{0ij}^-$ are normalisation constants required so that we can think of this as a mirror symmetric state discrimination problem. Explicitly,
\begin{align}
    \begin{aligned}
    \mathcal{N}_{001}^- &= \sqrt6,\\
    \mathcal{N}_{010}^- &= \sqrt{\frac{6(\alpha + 4\beta)}{4\beta + 3\alpha}} = \mathcal{N}_{011}^-.
    \end{aligned}
\end{align}

Now, in \cite{mirrorSym}, the states to be discriminated are written as
\begin{subequations}
\begin{alignat}{4}
    |\psi_{1}\rangle &=  |1\rangle,\\
    |\psi_{2}\rangle &= \cos \theta |1\rangle - \sin \theta |0\rangle,\\
    |\psi_{3}\rangle &= \cos \theta |1\rangle + \sin \theta |0\rangle,
\end{alignat}
\end{subequations}
such that $|\psi_{2,3}\rangle$ happen with probability $p_{2,3} = p$ and $|\psi_{1}\rangle$ with probability $p_1 = 1-2p$. So, we can let
\begin{align}\label{SinCos}
    \begin{aligned}
    \cos \theta &= \sqrt{\frac{\beta}{4\beta + 3\alpha}},\\
    \sin \theta &= \sqrt{\frac{3(\alpha + \beta)}{4\beta + 3\alpha}},
    \end{aligned}
\end{align}
and $q_{010}^-, q_{011}^- = p^-$, $q_{001}^- = 1-2p^-$ where
\begin{equation}\label{MirrorSymStatesProbs}
    q_{0ij}^- = P(\hat{\rho}_{0ij} | \hat{\pi}_-, s=1/2)
\end{equation}
is the probability of being in the state $|\psi_{0ij}^-\rangle$, and we have added (with respect to \cite{mirrorSym}) a superscript to $p$ to distinguish the two outcomes of the intermediate measurements.

So, using Eq.\,(\ref{MirrorSymStatesProbs}), we can find $p^- = q_{0ij}^-$. Using Bayes' theorem, we find that
\begin{equation}
    p^- = \frac{P\!\left(\hat{\pi}_-\otimes\mathbbm1, s = \frac12 \Big| \hat{\rho}_{0ij}\right)\!P(\hat{\rho}_{0ij})}{P\!\left(\hat{\pi}_-\otimes\mathbbm1, s = \frac12\right)}.
\end{equation}
Note that requiring $s = 1/2$ is equivalent to projecting the state $\hat{\rho}_{0ij}$ onto the $s=1/2$ space $(\mathcal{Q}_{\frac12} \otimes \mathcal{P}_{\frac12}) \otimes (\mathcal{Q}_{\frac12} \otimes \mathcal{P}_{\frac12})^*$. Denoting this projector by $\hat{P}_{\frac12}$,
\begin{widetext}
\begin{equation}
    P\!\left(\!\hat{\pi}_-\otimes\mathbbm1, s = \frac12 \Big| \hat{\rho}_{0ij}\! \right)\! = P\!\left(\!\hat{\pi}_-\otimes\mathbbm1, \hat{P}_{\frac12} \Big| \hat{\rho}_{0ij}\!\right)\!
    = \text{Tr}\!\left(\hat{P}_{\frac12} \!\left(\sqrt{\hat{\pi}_-}\otimes \mathbbm1\right)\! \hat{\rho}_{0ij} \!\left(\sqrt{\hat{\pi}_-}\otimes \mathbbm1\right)\! \hat{P}_{\frac12}\right)\!.
\end{equation}
\end{widetext}
The denominator can be found using
\begin{equation}
    P\!\left(\!\hat{\pi}_-\otimes\mathbbm1, s = \frac12\right)\! = \sum \limits_{ij} P(\hat{\pi}_- \otimes \mathbbm1, \hat{P}_{\frac12}| \hat{\rho}_{0ij})P(\hat{\rho}_{0ij}),
\end{equation}
from which it follows that
\begin{equation}
    p^- = \frac{3\alpha + 4\beta}{6(\alpha + 2\beta)}.
\end{equation}

Now, according to \cite{mirrorSym}, if
\begin{equation}
    p \geq \frac{1}{2+ \cos \theta (\cos \theta + \sin \theta)},
\end{equation}
$a = 1$. Else,
\begin{equation}
    a = \frac{p\cos \theta \sin \theta}{1 - p(2 + \cos^2 \theta)}.
\end{equation}
Substituting $p^-$ for $p$, $a_-$ for $a$ and our expressions for $\sin \theta, \cos \theta$ given in Eq.\,(\ref{SinCos}), this can be restated in the following way: if
\begin{equation}
    \alpha \geq 2\beta,
\end{equation}
$a_- = 1$. Else,
\begin{equation}
    a_- = \sqrt{\frac{\alpha + \beta}{3\beta}}.
\end{equation}
When coupled with the constraints on $\alpha,\beta$ given in Eq.\,(\ref{weakMeas1}), we obtain
\begin{equation}
    a_- = \begin{cases}
    \sqrt{\frac{\alpha + \beta}{3\beta}} &\text{ if } \alpha \in [0,\text{min}\{ 1-\beta, 2\beta\}],\\
    \quad ~ ~ 1 &\text{ if } \alpha \in (2\beta,1-\beta] \text{ with } 2\beta < 1- \beta 
    \end{cases}
\end{equation}
such that $\beta \in [0,1]$.

\providecommand{\noopsort}[1]{}\providecommand{\singleletter}[1]{#1}%
%


\end{document}